\documentclass[runningheads]{llncs}
\usepackage{graphicx}

\begin{document}

\title{EtrusChain:File Storage with DNA and Blockchain}

\author{Onur Yıldırım}

\authorrunning{Onur Yıldırım}

\maketitle

\begin{abstract}
This article proposes a blockchain-based file storage system that utilizes DNA encryption for enhanced security. The system utilizes blockchain technology to provide decentralized and tamper-proof file storage, while DNA encryption is employed to further strengthen data protection. The proposed system employs a unique approach to encryption by utilizing DNA sequences as keys, which enhances data security and privacy. Additionally, the use of blockchain technology ensures that all file storage and access operations are transparent, immutable, and distributed among a network of nodes, making it resistant to tampering and unauthorized access. The proposed system represents a significant advancement in file storage security and provides a foundation for future research and development in the field of blockchain-based data storage.

\end{abstract}

\section{Introduction}
\subsection{A Looking to Data Storage Systems}
In this era of ubiquitous data, the exigency for efficacious and secure file storage systems has become paramount. Existing solutions often fall short in terms of security, efficiency, and user agency. To redress these deficiencies, our team has conceived an innovative decentralized file storage system that amalgamates DNA encryption and blockchain technology. By fusing novel aspects of data encryption and storage, our system proffers unparalleled levels of security, confidentiality, and sovereignty over stored data. In this treatise, we explicate the architecture, pivotal features, and meritoriousness of our system, and demonstrate its potential to metamorphose the field of data storage.

In the era of the digital age, storing data has become a necessity, both for individuals and organizations. However, this has led to the problem of centralization in data storage, where data is stored in large centralized servers owned by a few entities, making it vulnerable to security breaches and data manipulation. These centralized solutions often lack transparency and are subject to censorship, surveillance, and control by the owners of the servers. This has led to the need for a decentralized file storage system that offers a secure and privacy-preserving solution.

Existing solutions to decentralized file storage, such as InterPlanetary File System (IPFS) and Sia, have limitations in their storage capacity, performance, and user-friendliness. They also lack the encryption required to protect the privacy and security of the data stored in the network. These limitations pose a challenge for users who require large storage capacity, high performance, and strong security measures.

In this paper, we present a novel decentralized file storage system that offers an effective, secure, and privacy-preserving alternative to centralized solutions. Our system leverages the use of DNA encryption and blockchain technology to ensure that data is stored in a secure and immutable manner. The use of DNA encryption makes it practically impossible for unauthorized parties to access the data, even if they gain access to the storage nodes.

Our system differs from existing solutions in its unique combination of DNA encryption and blockchain technology. This combination provides a higher level of security and privacy for users, ensuring that their data is safe from any unauthorized access or manipulation. Additionally, our system offers high performance and scalability, making it suitable for a wide range of applications.

Overall, the need for a decentralized file storage system has become increasingly important in today's digital world. Our system presents a more effective, secure, and privacy-preserving alternative to centralized solutions, addressing the limitations of existing decentralized file storage solutions. The combination of DNA encryption and blockchain technology offers a unique and innovative solution to the problem of centralized data storage, making it an ideal choice for users who value their data privacy and security

\subsection{Related Works}
\subsubsection{Google Drive}
Google Drive is a cloud-based file storage and synchronization service offered by Google. It allows users to store and share files and folders across multiple devices and collaborate with others in real-time. While Google Drive offers many features and benefits, it is centralized, and users must trust Google to protect their data from unauthorized access or tampering. Additionally, the data stored on Google Drive is not encrypted, which increases the risk of data breaches and privacy violations.

\subsubsection{Sia}
Sia is a decentralized file storage system that uses blockchain technology to create a network of nodes that store and retrieve files. It provides users with increased privacy and security by encrypting data before it is uploaded to the network. Sia also offers competitive pricing compared to centralized cloud storage services. However, the user experience can be challenging for some, and the system's performance can be affected by the number of nodes on the network

\subsubsection{Storj}
Storj is another decentralized file storage system that uses blockchain technology to create a network of nodes that store and retrieve files. Like Sia, Storj offers users enhanced privacy and security by encrypting data before it is uploaded to the network. Storj also uses a unique approach to data redundancy, which helps ensure that data is always available. However, like Sia, Storj can suffer from performance issues due to the number of nodes on the network.

\subsubsection{Filecoin}
Filecoin is a decentralized file storage system that utilizes a blockchain-based marketplace to incentivize users to store and retrieve files. Filecoin aims to create a more efficient and cost-effective way to store files by allowing users to purchase storage space from other users on the network. While Filecoin is still relatively new, it has the potential to revolutionize the way files are stored and distributed.

\subsubsection{Arweare}
Arweave is a decentralized file storage system that uses blockchain technology to create a permanent and tamper-proof record of all data stored on the network. Arweave uses a unique approach to file storage called the "permaweb," which ensures that files are always available and cannot be deleted or modified. Arweave is an exciting development in the field of decentralized file storage and has the potential to revolutionize the way data is stored and shared online.

\subsection{Comparison with Our System}
Google Drive, Sia, Storj, Filecoin, and Arweave are all examples of file storage systems with varying strengths and weaknesses. While Google Drive offers a user-friendly experience and real-time collaboration, it is centralized and lacks proper encryption for data security. Sia and Storj offer increased privacy and security through encryption, but their performance can suffer from the number of nodes on their networks. Filecoin's blockchain-based marketplace incentivizes users to store and retrieve files more efficiently, but it is still a new system that needs further development. Finally, Arweave's permaweb provides tamper-proof storage, but it can be limited in terms of accessibility and flexibility.

Our proposed blockchain-based file storage system, supported by DNA-encryption algorithms, addresses the limitations of these systems by providing a decentralized and secure storage solution that is also efficient and flexible. The use of blockchain technology ensures that data is stored and distributed across a network of nodes, eliminating the need for a centralized authority and minimizing the risk of data breaches and tampering. Additionally, the use of DNA-encryption algorithms ensures that data is properly encrypted and secure, even in the case of unauthorized access.

Overall, our system provides a robust and secure file storage solution that is both user-friendly and cost-effective. By leveraging the strengths of both blockchain and DNA-based systems, we offer a unique and innovative approach to file storage that addresses the limitations of existing systems.

\section{Decentralized Storage Systems}
\subsection{Introduction} 
In recent years, there has been a significant shift towards decentralization in various aspects of technology, including file storage. Decentralized systems refer to those in which there is no central authority controlling the flow of data or the maintenance of the network. Instead, nodes in the network have equal power, and decisions are made based on consensus.

Decentralized file storage systems have a number of advantages over centralized systems, including increased security, scalability, and fault tolerance. In a centralized system, all data is stored in one central location, making it vulnerable to hacking or other forms of attack. In contrast, in a decentralized system, data is distributed across multiple nodes, making it much more difficult for an attacker to compromise the system. Additionally, decentralized systems can be more scalable than centralized systems, as the workload can be distributed across multiple nodes. Finally, decentralized systems can be more fault-tolerant than centralized systems, as a failure in one node does not necessarily lead to a failure in the entire system.

The main logic or architecture for decentralized file storage involves dividing files into smaller pieces and distributing them across multiple nodes in the network. This process is known as sharding, and it enables files to be stored and retrieved more efficiently than in a centralized system. When a user uploads a file to the system, the file is broken up into smaller pieces, and these pieces are stored on different nodes in the network. When the user wants to retrieve the file, the system retrieves the different pieces from the nodes and reassembles them into the original file.

One example of a decentralized file storage system is IPFS (InterPlanetary File System), which uses a distributed hash table (DHT) to store and retrieve files. IPFS works by breaking files into smaller pieces called blocks, which are then given a unique identifier called a hash. The blocks are stored on different nodes in the network, and the hash allows the system to retrieve the blocks and reassemble them into the original file when needed.

\subsection{Working Principles of Blockchain-Based File Storage Systems}
Decentralized systems are a new way of organizing information that aims to eliminate the need for a central authority to control and validate transactions. Instead, they rely on a network of peers that collaborate to achieve consensus on the validity of each transaction.

One of the main advantages of decentralized systems is that they are more secure than centralized ones. Since there is no central authority that controls the system, it is much harder for attackers to hack into it and manipulate the data. Additionally, decentralized systems are more resistant to censorship and denial-of-service attacks since there is no central point of failure.

Disturbed systems are a type of decentralized system that relies on a network of nodes that are geographically distributed. Each node stores a copy of the database, and transactions are validated by a consensus mechanism that ensures that all nodes agree on the validity of each transaction. blockchain-based  file storage systems rely on distributed ledger technology to store and manage data. Unlike centralized systems, where data is stored on a single server or cluster of servers, blockchain-based  systems store data across a distributed network of nodes. This network is maintained and secured by a consensus mechanism that ensures the integrity and accuracy of the data stored on the network.

One of the key principles of blockchain-based file storage systems is the use of cryptographic algorithms to protect data. When data is uploaded to the network, it is encrypted using a public key. This means that only the owner of the private key can access the data. In addition, blockchain-based systems use hash functions to ensure that data cannot be tampered with or altered without detection.

Another important principle of blockchain-based file storage systems is the use of smart contracts to manage the storage and retrieval of data. Smart contracts are self-executing programs that run on the blockchain and can be programmed to automatically execute certain actions based on predefined conditions. In the context of file storage, smart contracts can be used to automatically pay storage providers for the amount of data stored or retrieved from the network.

Consensus mechanisms are also a crucial part of blockchain-based file storage systems. These mechanisms ensure that all nodes on the network agree on the state of the ledger and prevent malicious actors from tampering with the data stored on the network. Some of the most popular consensus mechanisms used in blockchain-based file storage systems include proof-of-work, proof-of-stake, and proof-of-replication.

Finally, blockchain-based file storage systems also rely on a decentralized architecture to ensure that data is distributed across multiple nodes on the network. This helps to prevent single points of failure and ensures that the network is more resilient to attacks or failures. In addition, a decentralized architecture helps to ensure that data is more accessible to users around the world, regardless of their location.

Overall, the working principles of blockchain-based file storage systems are designed to ensure that data is secure, accessible, and reliable. By leveraging cryptographic algorithms, smart contracts,  and decentralized architectures, these systems have the potential to revolutionize the way we store and manage data in the digital age.
\section{File Storage with DNA Systems}
\subsection{Introduction} 
DNA computing is a relatively new field that involves using DNA molecules to perform computations. In DNA computing, the four nucleotide bases that make up DNA (adenine, guanine, cytosine, and thymine) are used to represent binary code. This enables DNA molecules to store and process information in a way that is fundamentally different from traditional computing systems.

There is a growing interest in using DNA-based storage systems for file storage due to their potential for high density and long-term storage. DNA-based storage systems can potentially store vast amounts of data in a very small space, making them much more efficient than traditional storage systems. Additionally, DNA-based storage systems can potentially store data for thousands of years without degradation, making them ideal for long-term archival storage.

There are several examples of DNA-based storage systems, including a system developed by researchers at the European Bioinformatics Institute that was able to store 2.2 petabytes of data in a single gram of DNA. Another example is a system developed by researchers at Harvard University that was able to store an entire book in DNA.

The main logic or architecture for DNA-based file storage involves converting digital data into DNA code and then synthesizing the DNA molecules to store the data. When the data needs to be retrieved, the DNA molecules are sequenced and converted back into digital data.But we dont made this because this method is so expensive.We will try another solutions of DNA-based storage systems

\subsection{Working Principles of DNA-based File Storage Systems}
DNA-based file storage systems rely on the principle of encoding digital data into DNA sequences, which can then be stored and retrieved using DNA synthesis and sequencing technologies. \textbf{There are two main methods for encoding data into DNA: the Church-Gao-Kosuri (CGK) method and the DNA Fountain method.}

The CGK method involves converting digital data into binary code, which is then converted into DNA nucleotide sequences. The binary code is divided into smaller segments, which are then converted into nucleotide sequences using a one-to-one mapping system. These nucleotide sequences are then synthesized into DNA strands, which can be stored in a DNA database or library. When the stored data is needed, the DNA strands can be sequenced and converted back into binary code for use.

\subsubsection{Example Algorithm of CGK Method}
\textbf{Algorithm}: Church-Gou-Kosouri method with DNA nucleotides

\textbf{Inputs}:
- s: DNA sequence
- M: integer

\textbf{Outputs}:
- p, q: non-trivial factors of n

1. \textbf{Define} f(x) = (x\^2) mod n + M

2. \textbf{Set} x = 2, y = 2, d = 1

3. \textbf{While} d = 1 do
  
   3.1 \textbf{Set} x = f(x)
  
   3.2 \textbf{Set} y = f(f(y))

   3.3 \textbf{Set} d = gcd(|x - y|, n)

4. \textbf{If} d = n then
   
   4.1 Go to step 2

5. \textbf{End if}

6. \textbf{Set} p = d, q = n/d

7. \textbf{Translate} DNA sequence into a number n
   
   7.1 \textbf{Replace} 'A' with 00
  
   7.2 \textbf{Replace} 'C' with 01
  
   7.3 \textbf{Replace} 'G' with 10
   
   7.4 \textbf{Replace} 'T' with 11
   
   7.5 \textbf{Convert} resulting binary string to a decimal number
  
   7.6 \textbf{Set} n = decimal number

8. \textbf{Translate} factors p and q into DNA sequences
   
   8.1 \textbf{Convert} p and q to binary strings
   
   8.2 \textbf{Divide} binary strings into groups of two digits
  
   8.3 \textbf{Replace} '00' with 'A'
  
   8.4 \textbf{Replace} '01' with 'C'
  
   8.5 \textbf{Replace} '10' with 'G'
   
   8.6 \textbf{Replace} '11' with 'T'
   8.7 Con\textbf{}catenate resulting nucleotides to form DNA sequences for p and q

9. \textbf{Return} DNA sequences for p and q

The DNA Fountain method, on the other hand, uses a different approach to encode data into DNA. This method involves dividing the digital data into small segments and encoding each segment into a unique DNA sequence. The DNA sequences are then merged together using a fountain coding system, which creates a continuous stream of DNA sequences. This stream can be synthesized into DNA strands and stored in a DNA library. To retrieve the data, the DNA strands are sequenced and decoded using the fountain coding system.

Both of these methods provide a\textbf{ highly secure and durable storage solution for digital data}. DNA molecules are stable and can remain intact for thousands of years, making them an ideal storage medium for long-term archiving. DNA-based storage systems also offer a high storage density, with the potential to store large amounts of data in a small physical space.

However, DNA-based storage systems also face several challenges, such as the high cost and time required for DNA synthesis and sequencing. Additionally, errors can occur during the synthesis and sequencing process, which can result in data loss or corruption. These challenges are being addressed through ongoing research and development in the field.

Overall, DNA-based file storage systems offer a promising solution for the long-term storage and archiving of digital data. With ongoing advancements in DNA synthesis and sequencing technologies, these systems have the potential to become a mainstream method for digital data storage in the future.

\section{Our System: EtrusChain}

\subsection{Introduction} 
Our proposed system is a blockchain-based file storage system that incorporates DNA computing to enable high-density, long-term storage. The system architecture involves dividing files into smaller pieces and storing them on different nodes in the network using a sharding mechanism similar to that used in decentralized file storage systems. The difference is that instead of storing the data on traditional storage devices, the data is converted DNA code and stored in DNA molecules, which are distributed across the nodes in the network.

The system uses a blockchain to maintain a secure and transparent ledger of all file transactions within the network. When a user uploads a file, the system creates a hash of the file and stores it on the blockchain. This ensures that the file is tamper-proof and can be easily retrieved by the user at any time. The system also incorporates smart contracts, which enable the creation of decentralized applications that can interact with the file storage system.

To store data in DNA molecules, the system first converts the digital data into DNA code using a process known as DNA synthesis. The synthesized DNA is then stored in small, glass beads, which are distributed across the nodes in the network. The DNA molecules are then sequenced and converted back into digital data when the data needs to be retrieved.

The use of DNA-based storage in our system has several advantages over traditional storage systems. DNA-based storage has the potential for much higher density, allowing for vast amounts of data to be stored in a very small space. Additionally, DNA-based storage can potentially store data for thousands of years without degradation, making it ideal for long-term archival storage.

\subsection{System Architecture}

The architecture of our system comprises three main components: the blockchain layer, the DNA computing layer, and the data models/API layer. Let's delve into each of these components in detail.\textbf{(Figure 1)}

\subsubsection{Blockchain Layer}

At the core of our system is the blockchain layer, which serves as the foundation for secure and transparent file storage. The blockchain acts as a distributed ledger that stores the metadata and cryptographic hashes of files, ensuring their integrity and immutability. The use of blockchain technology eliminates the need for a central authority or intermediary, providing a decentralized and tamper-proof system for storing and managing files. Our system utilizes a consensus mechanism, such as Proof of Work (PoW) or Proof of Stake (PoS), to validate transactions and ensure the integrity of the blockchain.

To ensure the security of the system, the blockchain layer consists of multiple nodes distributed across the network. Each node maintains a copy of the blockchain and participates in the consensus process to validate transactions. When a new transaction is initiated, it is broadcasted to the network for validation. The nodes then compete to solve a complex mathematical problem, with the winner being the first to solve it and add the transaction to the blockchain. This ensures that the blockchain remains tamper-proof and immutable, as any attempt to modify the blockchain would require the modification of all copies of the blockchain across the network.

\subsubsection{DNA Computing Layer}

The DNA computing layer is where the magic happens. DNA, which is known for its incredible storage capacity, is utilized as a means to store files in our system. DNA strands can be synthesized and manipulated to represent binary data, making them an ideal candidate for storing digital files. To store a file in our system, it is first converted into binary data, and then encoded into DNA strands using techniques such as DNA converting

In the DNA computing layer, there are several processes that take place to ensure the accuracy and integrity of the stored data. Firstly, the DNA strands representing the file are checked for errors using DNA sequencing techniques. This process ensures that the DNA strands representing the file are accurate and have not been corrupted. Secondly, the DNA strands are stored in a controlled environment, such as a laboratory, to prevent any external factors from affecting their stability. Finally, the location and access permissions to the DNA strands are recorded in the blockchain for transparency and accountability.

\subsubsection{Data Models/API Layer}
The data models and APIs in our system provide the interface for users to interact with the blockchain and DNA computing layers. The data models define the structure and format of the data stored in the blockchain, including file metadata, cryptographic hashes, and transaction information. The APIs allow users to interact with the system, such as uploading files, retrieving files, and managing permissions. Our system provides a user-friendly and secure interface for users to interact with the blockchain and DNA computing layers seamlessly.

The data models and APIs layer consists of several components that work together to provide a seamless user experience. Firstly, there is the file upload API, which allows users to upload files to the system. When a user uploads a file, it is converted into binary data and then encoded into DNA strands. The DNA strands are then stored in a secure environment, and the location and access permissions are recorded in the blockchain.

Secondly, there is the file retrieval API, which allows users to retrieve files from the system. When a user initiates a file retrieval request, the blockchain layer validates the permissions and retrieves the location of the DNA strands from the blockchain. The DNA strands are then decoded back into binary data using DNA sequencing techniques, and the binary data is reconstructed into the original file format.

Finally, the data models and APIs layer provide the user interface for interacting with the blockchain and DNA computing layers. The data models define the structure and format of the data stored in the blockchain, including file metadata, cryptographic hashes, and transaction information. The APIs allow users to interact with the system, such as uploading files, retrieving files, and managing permissions.
The DNA computing layer also employs various algorithms and protocols to ensure the security and integrity of the stored files. The DNA strands are synthesized and manipulated using techniques such as Church-Gao Kosouri method,DNA Fountain method,Sanger sequencing to encode and decode the binary data. These techniques are well-established in the field of molecular biology and have been used in various DNA computing applications.

The file storage and processing workflow in our system is designed to ensure the security, transparency, and efficiency of file storage and retrieval. The uploaded files are first converted into binary data and encoded into DNA-based algorithms. The location and access permissions to the DNA strands are recorded in the blockchain for transparency and accountability.

When a user wants to retrieve a file from our system, they initiate a request through the user interface or API. The request is processed by the blockchain, which validates the permissions and retrieves the location of the DNA strands from the blockchain. The DNA strands representing the file are then retrieved from the storage location and decoded back into binary data using DNA sequencing techniques. Finally, the binary data is reconstructed into the original file format, which is returned to the user for download or further processing.

In \textbf{result of these operations}, our system provides a secure, transparent, and efficient way to store and retrieve digital files using a combination of blockchain and DNA computing technologies. The use of these cutting-edge technologies ensures the integrity and longevity of the stored files, while the user-friendly interface allows for easy interaction with the system. The various algorithms and protocols employed in the system ensure the security and transparency of the entire process, making it a reliable and trustworthy solution for digital file storage.

\subsection{Layer Plan}

\subsubsection{Blockchain Layer}
1.A user initiates a transaction to upload a file to the system.

2.The transaction is broadcasted to the network of nodes running the blockchain software.

3.The nodes validate the transaction using the special consensus mechanism to ensure the transaction is legitimate.

4.The validated file/data transaction is then added to the blockchain as a new block, containing the metadata and cryptographic hash of the file.

5.The new block is added to the chain, which is distributed to all nodes in the network, ensuring the immutability and transparency of the stored files.

Note:We are examining the layered construction on \textbf{Figure 2}

\subsubsection{DNA Layer}

1.After the blockchain layer has validated and stored the metadata and cryptographic hash of the file, the file is converted into binary data.

2.The binary data is then encoded into DNA strands using techniques such as DNA synthesis, creating a unique DNA sequence for the file.

3.The DNA strands are then stored in a controlled and secure environment, such as a laboratory, to ensure their integrity and longevity.

4.The location and access permissions of the DNA strands are recorded in the blockchain for transparency and accountability.

5.When a user requests to retrieve a file, the blockchain layer validates their permissions and retrieves the location of the DNA strands from the blockchain.

6.The DNA strands are then retrieved from the storage location and decoded back into binary data using DNA sequencing algorithms.

We developed a plan for characterizing this layer on \textbf{Figure 3}

\subsubsection{Data Model/API Layer}

1.The data models define the structure and format of the data stored in the blockchain, including file metadata, cryptographic hashes, and transaction information.

2.The APIs provide the interface for users to interact with the blockchain and DNA computing layers, allowing users to upload, retrieve, and manage files.

 3.Users can interact with the system through the user interface or API, initiating transactions and retrieving files.

 4.The system validates user permissions through the blockchain layer before allowing access to files.

 5.The APIs also provide functionality for managing permissions, such as granting access to specific users or revoking access.

Note:We are examining the layered construction on \textbf{Figure 4}

\textbf{Result:} By combining these layers, our system provides a decentralized, secure, and transparent solution for storing and managing files and we are examining the layered construction
in \textbf{Figure 1}

 \subsection{Summarized Layer Plan}

\begin{enumerate}
    \item Receive file upload request from user
\item Authenticate and authorize user access
\item Encrypt file using Church-Gao-Kosuri method or DNA Fountain coding algorithms
\item Fragment file data into smaller DNA sequences
\item Synthesize DNA sequences using DNA synthesis algorithms
\item Store synthesized DNA sequences on blockchain nodes
\item Read DNA sequences using DNA sequencing algorithms
\item Store encrypted file metadata and transaction information on blockchain
\item Validate transaction using PoS consensus mechanism
\item Index and store file data using RESTful APIs
\item Allow user to download file using API request
\item Decrypt file using Church-Gao-Kosuri method or DNA Fountain coding
\item Reassemble DNA fragments to retrieve original file data
\item Provide file to user for download or viewing.
\end{enumerate}

\subsection{Algorithm}
\begin{enumerate}
\item User uploads a file to the system.

\item The smart contract system creates a record on the blockchain that includes the file's unique hash, the user's identity, and a timestamp.

\item The smart contract system initiates the DNA encoding process by sending the file content to the DNA-based storage system.

\item The DNA-based storage system synthesizes the file content into DNA code and stores the DNA molecules in small glass beads.

\item The DNA-based storage system distributes the glass beads across the network for redundancy and fault tolerance.

\item  The smart contract system updates the blockchain with the location of the DNA molecules.

\item When a user requests to download a file, the smart contract system retrieves the file hash from the blockchain.

\item The smart contract system coordinates the retrieval of the relevant DNA molecules from the DNA-based storage system.

\item The DNA molecules are sequenced and decoded back into digital data.

\item The smart contract system returns the file content to the user.
\end{enumerate}

Note:We are examining the layered construction on \textbf{Figure 5}
\section{Conclusion}

In conclusion, our system stands out from its competitors due to its unique and innovative approach to file storage and management. The combination of blockchain and DNA computing technology provides a secure and decentralized system for storing and managing files, ensuring their integrity and immutability.

The use of blockchain technology eliminates the need for a central authority or intermediary, making our system more transparent and cost-effective compared to traditional file storage systems. Additionally, the utilization of DNA computing technology enables us to store vast amounts of data in a compact and durable format, ensuring the longevity of the stored files.

Our system is not only secure and efficient, but it is also profitable. By utilizing a pay-per-use model, we ensure that our customers only pay for the services they use, making our system more cost-effective than traditional storage methods. Furthermore, the increasing demand for secure and efficient file storage and management systems, coupled with the unique approach of our system, presents a significant profit potential for our company.

In fact, according to recent market research, the global market for blockchain-based data storage solutions is projected to grow from USD 233 million in 2021 to USD 1.4 billion by 2026, representing a compound annual growth rate (CAGR) of 42.8\%. This growth is driven by the increasing need for secure and transparent data storage solutions in various industries.

Therefore, by offering a unique and innovative system that addresses the needs of customers across multiple industries, we are poised for success in the growing blockchain-based data storage market.

\bibliographystyle{IEEEtran}
\bibliography{plain}

\cite{benisi2020blockchain}
\cite{bornholt2016dna}
\cite{church2012next}
\cite{ju2014what}
\cite{li2017blockchain}
\cite{lim2021novel}
\cite{wang2018blockchain}
\cite{yazdi2015dna}

\section{Figures / Appendix}

\begin{figure}[htbp]
  \centering
  \includegraphics[scale=0.4]{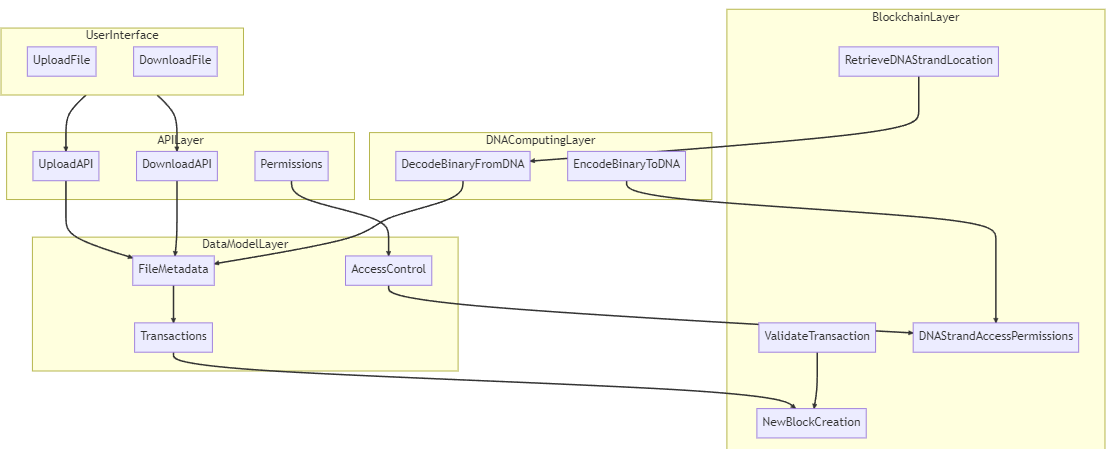}
  \caption{System Architecture Overview}
\end{figure}

\begin{figure}[htbp]
  \centering
  \includegraphics[scale=0.6]{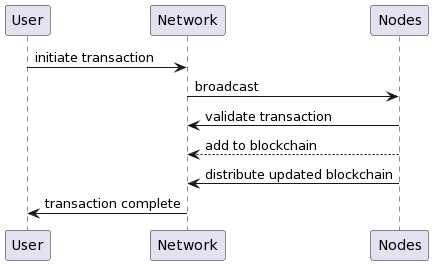}
  \caption{Overview of Blockchain Layer}
\end{figure}

\begin{figure}[htbp]
  \centering
  \includegraphics[width=\textwidth]{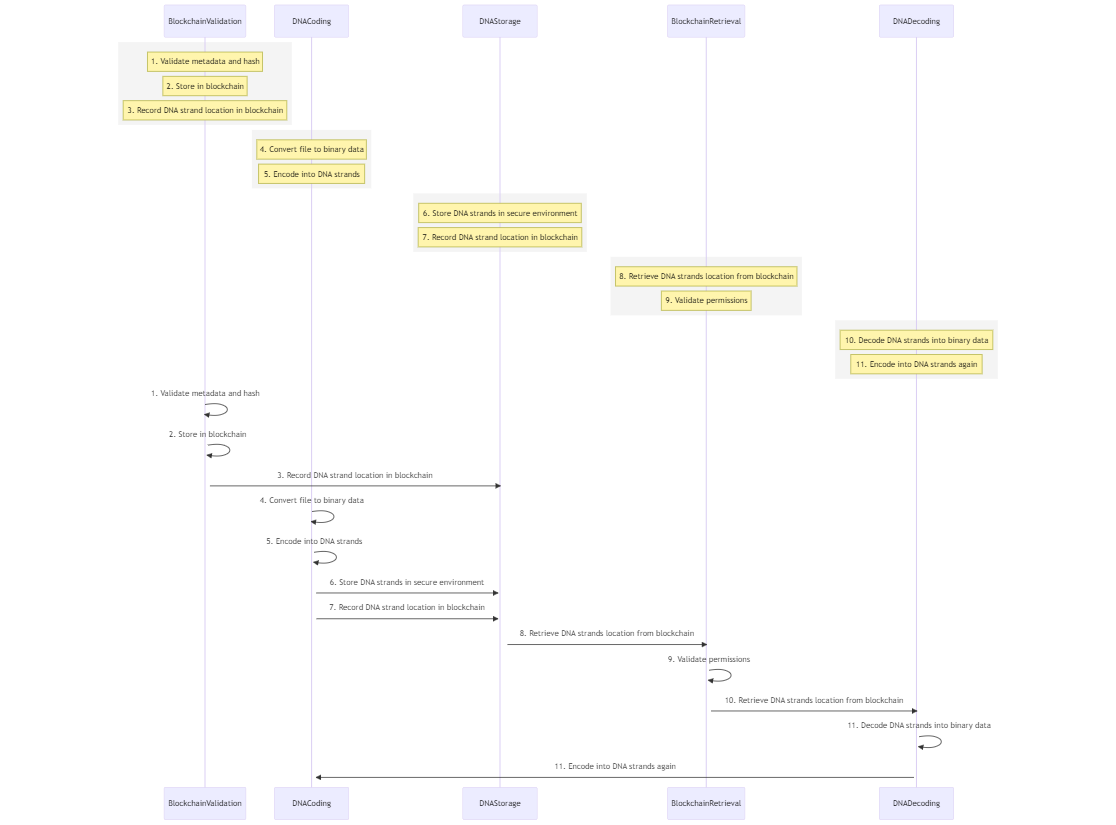}
  \caption{Overview of DNA-Layer}
\end{figure}

\begin{figure}[htbp]
  \centering
  \includegraphics[width=\textwidth]{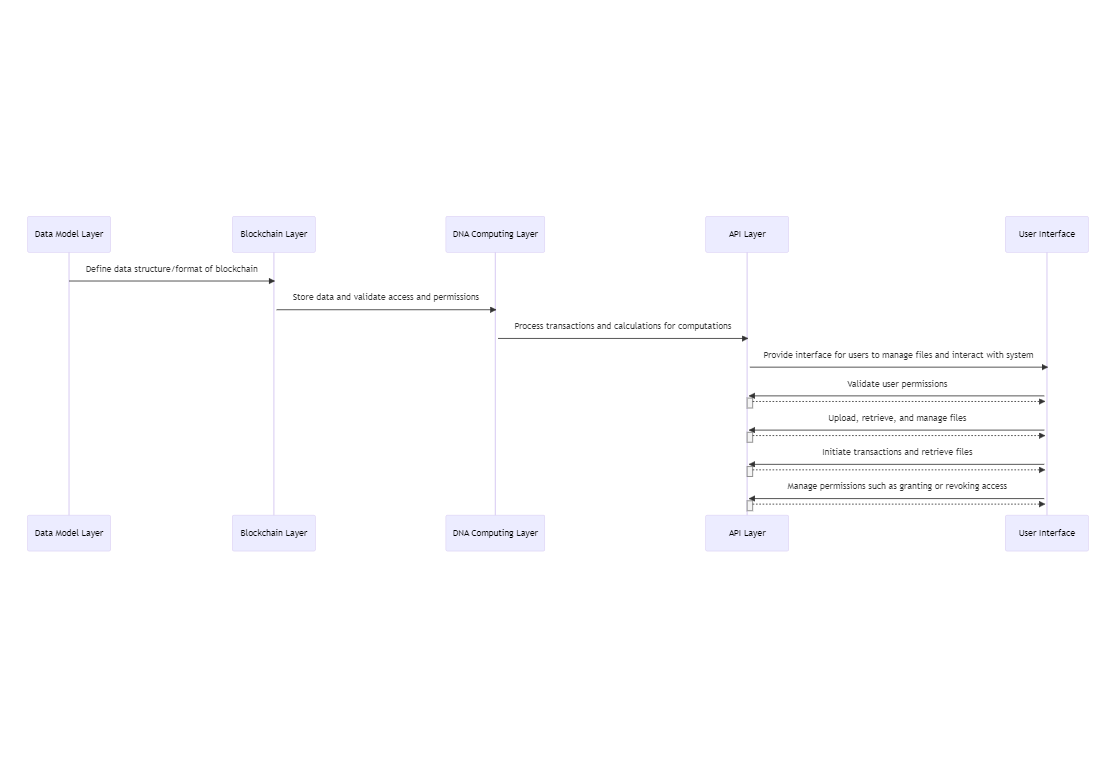}
  \caption{Overview of Data Model}
\end{figure}

\begin{figure}[htbp]
  \centering
  \includegraphics[scale=0.6]{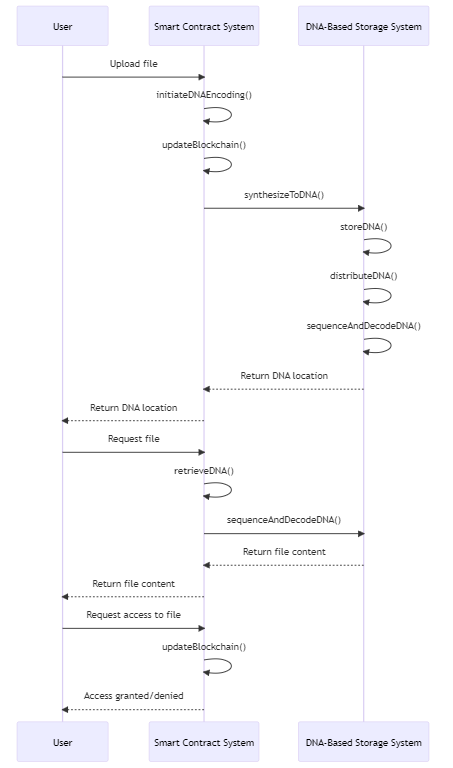}
  \caption{Algorithm Explained}
\end{figure}

\end{document}